\documentclass{article}
\usepackage{amssymb} % Useful mathematical tools
\usepackage{lipsum}
\usepackage{amsmath}
\usepackage{braket} % To write quantum states
\usepackage{float} % Omits 'float' error when captioning
\usepackage{caption} % Allows captioning
\usepackage{subcaption} % Allows subcaptioning
\usepackage{graphicx} % Required for inserting images
\usepackage{indentfirst} % To alleviate my OCD

\title{Quantum Neural Network Training of a Repeater Node}
\author{D. Fuentealba, J. Dahn, J. Steck, and E. Behrman}
\date{August 7, 2024}

\begin{document}

\maketitle

\section*{Abstract}
\textit{The construction of robust and scalable quantum gates is a uniquely hard problem in the field of quantum computing. Real-world quantum computers suffer from many forms of noise, characterized by the decoherence and relaxation times of a quantum circuit, which make it very hard to construct efficient quantum algorithms. One example is a quantum repeater node, a circuit that swaps the states of two entangled input and output qubits. Robust quantum repeaters are a necessary building block of long-distance quantum networks. A solution exists for this problem, known as a swap gate, but its noise tolerance is poor. Machine learning may hold the key to efficient and robust quantum algorithm design, as demonstrated by its ability to learn to control other noisy and highly nonlinear systems. Here, a quantum neural network (QNN) is constructed to perform the swap operation and compare a trained QNN solution to the standard swap gate. The system of qubits and QNN is constructed in MATLAB and trained under ideal conditions before noise is artificially added to the system to test robustness. We find that the QNN easily generalizes for two qubits and can be scaled up to more qubits without additional training. We also find that as the number of qubits increases, the noise tolerance increases with it, meaning a sufficiently large system can produce extremely noise-tolerant results. This begins to explore the ability of neural networks to construct those robust systems.}

\section*{Introduction}
One of the largest issues in current quantum computing research is the development of algorithms that fully leverage quantum phenomenon to produce a computational advantage \cite{08_paper}. Efficient and physically attainable algorithms for quantum computing have proven difficult to design. Our approach to this problem is to eliminate the need for humans to design the algorithm and allow a special neural network, called a quantum neural network (QNN), to construct the algorithm. Quantum neural networks are capable of producing robust and scalable control algorithms for quantum computers \cite{nam_paper}\cite{nathan_paper}. This project aims to demonstrate that these properties transfer well to control of a quantum repeater node, a linear operation consisting of swapping the states of a set of input qubits and a corresponding set of output qubits. Training a repeater node was chosen to demonstrate the effectiveness of the QNN approach because a repeater node has a well-known structure consisting of an n-qubit SWAP gate. Additionally, a quantum repeater does not have a measurement operation, meaning the system is linear. To construct the QNN, we first construct both the Hamiltonian acting on the system and the density matrix being acted upon. Next, a training method is devised using a Levenberg-Marquardt learning algorithm. To begin the actual training, a system of two qubits is trained and the parameters are copied to four, six, and eight qubits. Finally, various types of noise are introduced at different levels to test the robustness of the system. 

\section*{Literature Review} 
One of the most useful applications of a neural network can be seen when control is needed for a system with an unknown mathematical model \cite{Fund_of_NN}. Quantum control algorithms fit well within this category of problems. While the Hamiltonian of a set of qubits is, in a general sense, simple to calculate, manipulation of that Hamiltonian to produce a desired output within the physical limitations of the system is a difficult task. Using a neural network to devise the control scheme passes the difficulty to the network, and the new task is then to devise the proper architecture for the network and provide a suitable training set for it. In general, both of these tasks are less difficult than devising the quantum control algorithm. 

The structure of a quantum neural network is a hybrid system: a classical neural net and a quantum processor \cite{QML_Review}. In this approach, the quantum processor is queried by the neural net \cite{Hybrid}. The neural net sends parameters to the quantum system, the quantum system evolves in time given those parameters, and makes a measurement. The measurement is then passed back to the neural network, which calculates new parameters to send back to the quantum system. This iterative procedure is the same technique neural nets use to learn in any application. The probabilistic nature of quantum measurement is one of the barriers to algorithm construction, which neural networks are not immune to. To demonstrate this, it is convenient to define the neural net as a time-dependent Hamiltonian $H(t)$ acting on the quantum system, where the same neural net can act on classical bits. Then we have 
\begin{align}
|\psi_{f}\rangle=H(t)|\psi_{i}\rangle \\
\vec{s}_{f} = H(t)\vec{s}_{i}
\end{align}
but in general 
\begin{align}
|\psi_{f}\rangle=\alpha|0\rangle+\beta|1\rangle
\end{align}
so the quantum processor could measure $|0\rangle$ with probability $|\alpha|^{2}$ or $|1\rangle$ with probability $|\beta|^{2}$, meaning the neural network could get two different outputs for a single input without changing any weights. This presents problems in training the neural net, but with careful selection of initial conditions and training parameters, the net can still generalize with relative ease. However, quantum repeaters do not have a measurement operation, and their time evolution is a strictly linear set of differential equations. To be tested on actual hardware, a measurement operation will need to be added to determine if the algorithm is accurate. The true uniqueness of quantum systems lies in the vastness of the Hilbert space spanned by the entangled qubits. This Hilbert space is much larger than any spanned by any classical system. In terms of control, there is a much larger continuum of states that the system could take compared to a classical system. 

Another problem in current quantum technology is noise \cite{nam_paper}. Noise is generated by the environment around the quantum system such as heat or external EM fields. These external fields can add energy to the fragile quantum state \cite{QR_Imp}. When using neutral atoms, trapped ions, or other similar atomic qubits, noise can also be generated by spontaneous emission. Spontaneous emission can knock the qubit back into its ground state, and release a photon in the process, further damaging the signal. The rate of spontaneous emission is dictated by the electron transitions that comprise the ground and excited states. While the rate of noise production can be modeled adequately with information about the environment the qubit is in, the actual noise is very hard to predict, if not random. The noise frequently damages the outgoing signals from a quantum system, and it is generally accepted that near-term quantum computers will need to be able to operate within a noise threshold instead of being able to completely eliminate signal noise, a so-called "error-corrected quantum computer". 

This need for tolerance to noise and decoherence, referred to as the robustness of the system, is a property shown to be well suited to neural network control \cite{QML_Review}. While there are classical control techniques for noise handling, they often require the noise to be accounted for in the algorithm design. Noise in quantum systems can vary widely based on the system setup and thus could require many variations of a single algorithm for different noise scenarios. In contrast, when implemented to a particular system, it is easy for a neural network to be adjusted to fit the noise present via training on real data from the system. A neural net is then able to tune to virtually any system with almost negligible computational cost compared to a classical control approach \cite{Benchmarking_Paper}. The decoherence of the system is a uniquely quantum problem that causes the mixing of the states of the qubits, and quantum processors are benchmarked with a decoherence time. Operating within the decoherence time of the quantum system is then one of the main constraints of the control system. 

\subsection*{Quantum Repeater Nodes}
The purpose of this research project is to test the ability of neural networks to construct robust, scalable quantum control algorithms, and to test the robustness using a system with a well-known solution. A repeater node, therefore, provides a great proof of concept because it is both a well-understood system to construct and because of its linearity. 

A quantum repeater can be constructed as a system that initializes output qubits $|B_{1}\rangle,|B_{2}\rangle,...,|B_{n}\rangle$  and receives input qubits $|A_{1}\rangle,|A_{2}\rangle,...,|A_{n}\rangle$ \cite{Toward_QN}. The repeater then swaps each input state $|A_{i}\rangle$ with its corresponding state $|B_{i}\rangle$, where $|B_{i}\rangle$ is output from the system and $|A_{i}\rangle$ is stored in the repeater and remains entangled with $|B_{i}\rangle$. The repeater essentially functions as a signal amplifier, however, because the incoming signal cannot be copied, so the input and output signals are swapped instead. When the outgoing qubit is measured, the input signal collapses with the measurement. The output signal initialized inside the repeater can be arbitrary, and thus can generally be initialized to a standard state that is easy to produce.

The operations of a repeater are fully contained in a swap gate. A swap gate can be constructed out of a finite number of CNOT and Hadamard gates. Therefore, the swap gate of a repeater can be represented as a single unitary operation. This single operation is what is trained by the neural network in this project. In practice, the operation being trained will be some combination of laser pulses acting on the qubits. With the application of quantum communication in mind, a typical repeater node would be most likely constructed with the incoming and outgoing qubits as photons through fiber optic cable or in the open atmosphere \cite{Satellite_Paper}. How to create storage qubits inside the repeater, as well as how to capture the incoming photon are active research areas. Several of the leading theories on how to do this involve "slow light", a phenomenon that uses a control light to effectively slow down the group velocity of the incoming signal \cite{Toward_QN}. Once the signal is captured, it will then be transferred to a stationary object such as a trapped ion, neutral atom, or quantum dot. The output qubit will then be formed with a laser pulse which will emit a photon entangled with the stationary qubit. The original signal is then stored in the repeater with the output qubit emitted and entangled to the stationary qubit, accomplishing the goal of the repeater. 

Quantum repeaters also have a unique property in quantum systems: they're completely linear. The nonlinearity in quantum systems comes from a measurement operation, which will collapse the system into one of its eigenstates, in a probabilistic manner. However, in the swap operation, there is no measurement being made. In the case of training a neural net to do this operation, it removes a layer of complexity by removing the non-linearity. This also has the added benefit of allowing the net to be trained on an arbitrary basis, with an easy choice being the charge basis. 

\subsection*{Applications}
One eventual goal of quantum computing is the construction of one or several quantum nets. A quantum net is analogous to the classical internet, providing the ability to link several nodes of quantum computers and distribute entangled qubits. The distribution of entangled qubits is a central requirement in several quantum algorithms such as quantum key distribution and quantum teleportation. 

Quantum key distribution (QKD) is a very promising framework in cryptography, leveraging the qubits to establish a secure communication channel \cite{Mike_n_Ike}. For standard encryption/decryption practices, a key is needed to encode and decode the message being sent. Key distribution is also one of the most vulnerable steps in encryption, because anyone who gets a hold of the key is able to decrypt the information. However, key distribution is also the most necessary step. The question then is how to get the key from the sender to the receiver, and the answer is there is no classical way to distribute the key without some security risk. QKD provides a way to ensure that the classical communication channel used to distribute the key is actually secure. The most popular QKD algorithm is the BB84 protocol, which requires the distribution of entangled qubits from source to target, which are possibly thousands of miles apart. Imagine a sender, Alice, wants to send a secret message to her lab partner Bob while avoiding eavesdropping by their nemesis Eve. To make sure their communications are actually secure, Alice sends Bob a string of \textit{n} qubits over a public channel, with each qubit randomly encoded in either the $|0\rangle,|1\rangle$ or $|+\rangle,|-\rangle$ bases, which are not mutually orthogonal. Bob then measures all the qubits randomly along one of the basis, without knowing which one Alice encoded each qubit in. Alice then announces publicly her encoding scheme, and Bob throws out any qubits that he measured in a different basis than Alice encoded them in. What is then left are \textit{n'} qubits that are measured in the same basis they were encoded in, with 100\% certainty of being measured as it was encoded (no qubits are in superposition states). Then any qubit measurement that does not match how Alice encoded it is evidence that Eve was able to intercept the message and measured the qubits before Bob was able to! Eve could in theory get lucky and guess the correct basis for every qubit, but when \textit{n} is on the order of dozens or hundreds, her chances become $.5^{n}$, which approach 0 very rapidly. Of course, this is the ideal scenario, where noise is not accounted for. In reality, the qubits will be sent via fiber optic cable or through a satellite array, over long distances which will incur noise within the system. This step is where the importance of high-fidelity repeaters is highlighted. Noise in the system will flip or lose qubits, creating the need to establish a threshold for mismatched qubits, as opposed to being able to use a single mismatched qubit to determine if the channel is secure. Noiseless quantum communication is not realistic, but the noise needs to be small enough that Eve could not "hide" in it. Without high-fidelity repeaters, able to handle hundreds or thousands of qubits in a short time, long-distance qubit distribution becomes almost impossible. 

Similar to QKD, quantum teleportation is a very useful protocol that has the potential to be realized with near-term quantum computers \cite{Mike_n_Ike}. Even more so than QKD, quantum teleportation also requires extremely high-fidelity transmission of qubits over long distances. Image again Alice has produced the highly exotic quantum state $|\phi\rangle = \alpha|0\rangle+\beta|1\rangle$ in her lab and wants to send it to Bob, who needs it for his experiment. While their labs are a continent apart, they luckily already each have one half of an EPR state $|\psi_{a}\rangle$ and $|\psi_{b}\rangle$ distributed through their secure quantum network. To send $|\phi\rangle$ to Bob, Alice first passes $|\phi\rangle$ and $|\psi_{a}\rangle$ through a CNOT gate with $|\psi_{a}\rangle$ as the control qubit, and then passes $|\psi_{a}\rangle$ through a Hadamard gate. These operations produce the state 
\begin{flalign*}
|\Psi\rangle = & \frac{1}{2}|00\rangle\left(\alpha|0\rangle+\beta|1\rangle\right)+\frac{1}{2}|01\rangle\left(\alpha|1\rangle+\beta|0\rangle\right)\\ &+\frac{1}{2}|10\rangle\left(\alpha|0\rangle-\beta|1\rangle\right)+\frac{1}{2}|11\rangle\left(\alpha|1\rangle+\beta|0\rangle\right)
\end{flalign*}
where the two leftmost qubits in each term are the entangled pair $|\psi_{a}\rangle$ and $|\phi\rangle$, and the right term is Bob's qubit $|\psi_{b}\rangle$. Alice can then measure her two qubits, and tell Bob over a classical communication channel what her measurement was. With this information Bob can then perform a rotation about the X or Z axis and recover the original state $|\phi\rangle$. 

While this application on its face seems to be a way to distribute qubits without physically sending qubits, the process is actually reliant on a robust quantum network to distribute the "sacrificial" EDR pair needed to teleport the desired qubit. The EDR pair is destroyed as soon as the qubits are measured, so for each state that is transported, another pair of entangled qubits must be distributed.

\section*{Methodology}
To construct the quantum repeater system, we first construct the form of the Hamiltonian acting on the system as 
\begin{align}
\displaystyle H = \sum_{i=1}^n k_{i}\sigma_{x,i} + \sum_{i=1}^n \epsilon_{i}\sigma_{z,i}+\frac{1}{2}\sum_{i\ne j}^n \zeta_{i,j}\sigma_{(z,i),(z,j)} \nonumber
\end{align}
where $\sigma_{x}\;and\;\sigma_{z}$ are the Pauli spin matrices, and $k_{i},\;\epsilon_{i},$ and $\zeta_{i,j}$ are the tunneling, bias, and coupling coefficients for each qubit respectively (or qubit pair in the case of $\zeta$). The qubits being acted on are represented in their density matrix form $\rho=|\Psi\rangle\langle\Psi|$ and $|\Psi\rangle=|\psi_{n}\rangle\otimes\ldots\otimes|\psi_{2}\rangle\otimes|\psi_{1}\rangle$. The system then evolves in time according to the Schr\"odinger equation, $ i\hbar\frac{\partial \rho}{\partial t}=[H,\rho]$. This representation of the system has proven complimentary to the training of a QNN \cite{08_paper}\cite{nam_paper} because of the similarities between the quantum system and a neural network, where $|\rho_{i}\rangle$ and $|\rho_{f}\rangle$ are the input and output states respectively, and the vectors of $k,\epsilon,$ and $\zeta$ are the weights of the network. Additionally, instead of directly training $k,\;\epsilon,$ and $\zeta$, each was broken apart into Fourier coefficients with a finite number of frequencies for each parameter. At each time step, the parameters were calculated using the Fourier sum, which gave discrete time functions of the parameters for use in the Hamiltonian. 

Once the Hamiltonian %as a function of time is <- this seems redundant?
is constructed, the system can be evolved through time using computational methods, and a measure of error collected at $t_{f}$. For this system, a set time step fourth order Runge-Kutta method is used. The measure of error was set as the Frobenius norm of the difference of the target state and final state density matrices.

%The measure was set as the fidelity of the final state $\rho_{f}$ with the target state $|\psi_{t}\rangle$ for each pair. Explicitly,
%\begin{flalign}
%(fidelity)_{i} &=|\psi_{f}\rangle\rho_{f}\langle\psi_{t}| \\
%(error)_{i} &= 1-(fidelity)_{i} \\
%E &= \sqrt{\sum_{i=1}^n (error_{i})^2}
%\end{flalign}
The error E is then fed into a Levenberg-Marquardt weight update rule for each training epoch. 

With the framework of the system set, we then divided the simulation into two cases. The ideal case with no noise or decoherence, and the realistic case, where noise and decoherence are introduced. These were introduced into the density matrix during time evolution. At each timestep, a noisy density matrix $\rho_{noise}$ was constructed by generating a hermitian matrix of random values, adding it to the original density matrix $\rho$, and normalizing. The matrix of random values was constructed using either real numbers, pertaining to pure noise (random magnitudes), or complex numbers, pertaining to decoherence (random phases) \cite{nam_paper}. This allowed us to analyze the effects of noise and decoherence either separately, or, together (as would realistically occur). We could also adjust the strength of the noise and decoherence by multiplying the random values by a parameter we called "Rho Noise Power" (RNP). Because this simulation is being done at a general level, the noise values are somewhat arbitrary and are meant to demonstrate the overall trend of the QNN's robustness. The actual noise in a quantum circuit will of course depend on the properties of that specific circuit. This is discussed in Section 6. 

To train the QNN, we first trained the noiseless two-qubit case, that is, one incoming qubit \textit{A} being swapped with one output qubit \textit{B}. Because there is no measurement operation in the repeater, any state can be written as a linear combination of basis states, so training on a basis set such as the charge basis was able to generalize the net. With the two-qubit case trained, the parameters could then be copied in pairs, with all input qubits having the same parameters and all output qubits having the same parameters. For simplicity in adding qubits to the system, we chose to create pairwise input and output qubits so that \textit{A} and \textit{B} swap, \textit{C} and \textit{D} swap, and so on. Finally, noise was added to the system, and each system of qubits was retrained, with noise added, in order to make the systems account for noise. As will be discussed in Section 5, we found that increasing the number of qubits decreased the effect the noise had on the system. 

\section*{Results}
To first establish a baseline for training effectiveness, a set of two qubits was trained to sufficient error E, and those parameters were then transferred to the four-, six-, and eight-qubit cases with no additional training. The baseline results are summarized in Table \ref{baseline_rms} for a testing set consisting of the charge states and 70 randomly generated quantum states. A jump is seen in the RMS error when scaling up from 2 to 4 qubits, but stabilizes from 4 to 6 and 6 to 8 qubits. 

Next, noise was added to the system to test the network's resiliency. Tests were conducted with pure noise and decoherence separate, as well as together (called complex noise), all at various levels. We find that the network's resilience is excellent for RNPs of magnitude $10^{-4}$ and lower, as shown by the figures and tables. For noise levels of magnitude $10^{-6}$ and $10^{-5}$, the system is much more immune to complex noise in the 8 qubit case than previous cases. So while the network experiences a slight spike in error in the initial bootstrapping, it recovers as the system continues to get bigger. This is important, as at higher system sizes, it will become unnecessary to train the network further, which would be expensive and time-consuming.

% With a baseline for training effectiveness and resilience to noise established, attention now shifts to improving the results by training with noise in the system. 

\begin{table}[H]
\centering
\begin{tabular}{|c|c|}
\hline
$n_{qubits}$ & Trained RMS \\
\hline
2 & $2.1518 \times 10^{-7}$ \\ 
\hline
4 & $1.4430 \times 10^{-4}$ \\
\hline
6 & $5.1553 \times 10^{-5}$ \\
\hline
8 & $1.8730 \times 10^{-5}$ \\
\hline
\end{tabular}
\caption{Noiseless RMS Values}
\label{baseline_rms}
\end{table}

\begin{table}[H]
\centering     
\begin{tabular}{|c|c|c|c|}
\hline
RNP & Pure Noise RMS & Decoherence RMS & Complex Noise RMS \\
\hline
5e-6 & 0.0001 & 0.0008 & 0.0078  \\ 
\hline
5e-5 & 0.0002 & 0.0020 & 0.0188 \\
\hline
5e-4 & 0.0004 & 0.0038 & 0.0340 \\
\hline
\end{tabular}
\caption{Error Induced by Noise for 2 qubits}
\label{4qubit_RMS}
\end{table}

\begin{table}[H]
\centering     
\begin{tabular}{|c|c|c|c|}
\hline
RNP & Pure Noise RMS & Decoherence RMS & Complex Noise RMS \\
\hline
5e-6 & 0.0002 & 0.0017 & 0.0134  \\ 
\hline
5e-5 & 0.0004 & 0.0032 & 0.0198 \\
\hline
5e-4 & 0.0013 & 0.0106 & 0.0334 \\
\hline
\end{tabular}
\caption{Error Induced by Noise for 4 qubits}
\label{4qubit_RMS}
\end{table}

\begin{table}[H]
\centering     
\begin{tabular}{|c|c|c|c|}
\hline
RNP & Pure Noise RMS & Decoherence RMS & Complex Noise RMS \\
\hline
5e-6 & 0.0007 & 0.0046 & 0.0111  \\ 
\hline
5e-5 & 0.0012 & 0.0068 & 0.0112 \\
\hline
5e-4 & 0.0062 & 0.0111 & 0.0128 \\
\hline
\end{tabular}
\caption{Error Induced by Noise for 6 qubits}
\label{4qubit_RMS}
\end{table}

\begin{table}[H]
\centering     
\begin{tabular}{|c|c|c|c|}
\hline
RNP & Pure Noise RMS & Decoherence RMS & Complex Noise RMS \\
\hline
5e-6 & 0.0017 & 0.0034 & 0.0063  \\ 
\hline
5e-5 & 0.0024 & 0.0035 & 0.0035 \\
\hline
5e-4 & 0.0035 & 0.0036 & 0.0116 \\
\hline
\end{tabular}
\caption{Error Induced by Noise for 8 qubits}
\label{4qubit_RMS}
\end{table}

\begin{figure}[H]
\centering
\includegraphics[width=.7\textwidth]{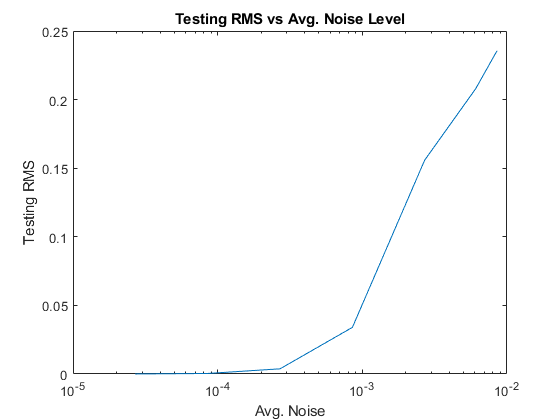}
\caption{RMS vs Average Noise Level for 2 qubits with Complex Noise.}
\end{figure}

\begin{figure}[H]
\centering
\includegraphics[width=.7\textwidth]{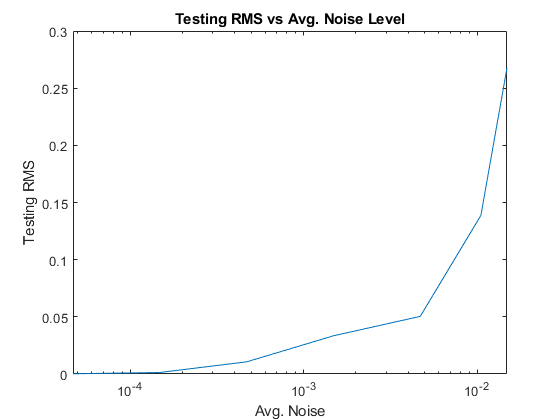}
\caption{RMS vs Average Noise Level for 4 qubits with Complex Noise.}
\end{figure}

\begin{figure}[H]
\centering
\includegraphics[width=.7\textwidth]{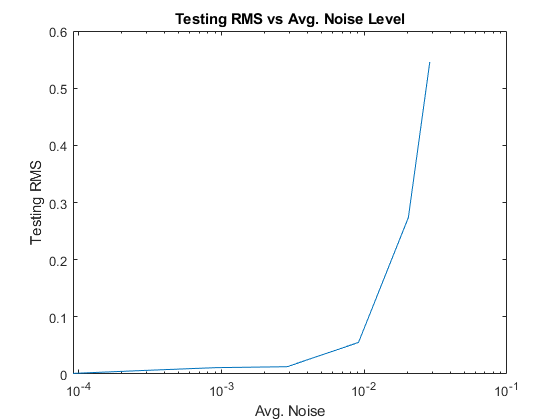}
\caption{RMS vs Average Noise Level for 6 qubits with Complex Noise.}
\end{figure}

\begin{figure}[H]
\centering
\includegraphics[width=.7\textwidth]{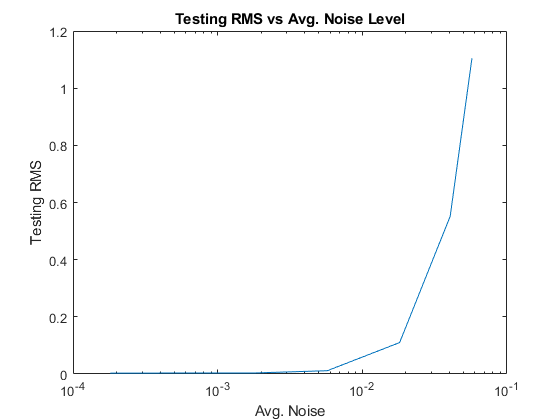}
\caption{RMS vs Average Noise Level for 8 qubits with Complex Noise.}
\end{figure}

\section*{Conclusion and Further Discussion}
The QNN was able to demonstrate both scalability and robustness. However, the system we created does not correlate directly with any specific hardware and must be put into the context of a demonstration of the methodology. 

As previously discussed in the paper, this simulation is offline learning. To be able to backpropagate the error through time, we have to know the state of the system at each timestep. This is clearly not possible, as any measurement of the system would cause it to collapse. Thus real hardware implementation would not be able to use this type of learning. Other control problems have run into this issue with classical systems. A common approach is to construct a model of the system for offline learning and perform all of the training in simulation. Once the model is sufficiently trained, the parameters can be transferred to the online system and if needed and further trained by some method not needing access to the state of the system at intermediate times. To construct the best possible model of the system for offline learning, many hardware-specific parameters must be known. Examples would be internal biases, relaxation and decoherence times, and actuation parameters. This was something we chose not to do in this project to produce a more general result. However, the logical progression of this research would be the implementation of this approach on actual hardware, ideally on a large variety of hardware. 

Even without hardware testing, these results are promising for the problem of the quantum repeater. Quantum repeaters will be the backbone of quantum networks and will need to be able to process large numbers of qubits with high fidelity. The QNN presented here allows the ability to trivially copy itself ad infinitum to handle the desired number of qubits. Additionally, with this scheme, it was also shown noise can be dealt with purely by the coupling of the qubits, meaning the QNN can be generalized by the training of a small number of qubits, and error correction can be handled by training the coupling coefficients.

\bibliographystyle{plain}
\bibliography{Manuscript}
\end{document}